
\magnification 1200
\hsize = 15 truecm \vsize = 22 truecm \hoffset = 1.0 truecm
\font\petit=cmr8
\def\m{M_j}
\def\ms{M_j^*}
\def\coset{{\bf C}(N_+/H_+)}
\def\Y{N_+/H_+}
\def\b{B^*(\widehat{sl}_2)}
\def\bj{B^j(\widehat{sl}_2)}
\def\n{{\bf n}}
\def\g{\widehat{sl}_2}
\def\a{{\bf a}}
\def\h{{\bf h}}
\def\half{{1\over2}}
\def\F{F^*(\widehat{sl}_2)}
\def\f{F (\beta)}
\def\BC{{\bf C}}
\def\l{j}
\def\+{e^{\int^z\,dt\,\beta(t) \gamma(t)}}
\def\sl{\widehat{sl}_2}

\def\j{J}
\def\-{e^{-\int^z\,dt\,\beta(t) \gamma(t)}}
\def\d{\partial}
\def\<{\langle}
\def\>{\rangle}
\rightline{LANDAU-94-TMP-8}
\rightline{submited to Mod. Phys. Lett. A}
\rightline{Dec 1994}
\vskip 3 truecm
\centerline{\bf GEOMETRICAL DESCRIPTION OF THE LOCAL INTEGRALS}
\centerline{\bf OF MOTION OF MAXWELL-BLOCH EQUATION}
\bigskip
\centerline{A. V. Antonov,%
\footnote{$^*$}%
{Supported in part by International Science Foundation (Grant M6N000)}
\footnote{$^1$}{E-mail: antonov@landau.ac.ru}
}
\centerline{\it Landau Institute for Theoretical Physics,}
\centerline{\it Kosygina 2, GSP-1, 117940 Moscow V-334, Russia}
\bigskip
\centerline{A. A. Belov,%
\footnote{$^{**}$}%
{Supported in part by International Science Foundation (Grant N89000)}
\footnote{$^2$}{E-mail: mitpan00@iki3.bitnet, subject: belov or
mitpan@oka.mitp.free.net, subject: belov}
}
\centerline{\it International Institute for Earthquake Prediction Theory}
\centerline{\it and Mathematical Geophysics,}
\centerline{\it Warshavskoe sh., 79, k. 2, Moscow 113556, Russia}
\bigskip
\centerline{B. L. Feigin,$^*$%
}
\centerline{\it Landau Institute for Theoretical Physics,}
\centerline{\it Kosygina 2, GSP-1, 117940 Moscow V-334, Russia}
\bigskip
{\petit We represent a classical Maxwell-Bloch equation and related to it
positive part of the AKNS hierarchy
in geometrical terms. The Maxwell-Bloch evolution is given by an infinitesimal
action of a nilpotent subalgebra $\n_+$ of affine Lie algebra $\sl$
on a Maxwell-Bloch phase space treated as
a homogeneous space of $\n_+$.
A space of local integrals of motion is described using cohomology methods.
We show that hamiltonian flows associated to the Maxwell-Bloch local
integrals of
motion (i.e. positive AKNS flows) are identified with an infinitesimal action
of an abelian subalgebra of the nilpotent subalgebra $\n_+$
on a Maxwell- Bloch phase space. Possibilities of
quantization and latticization of Maxwell-Bloch equation are discussed.}

\vfill\eject
\noindent
{\bf 1. Introduction}
\medskip\noindent
A powerful cohomological technique was proposed by B.~Feigin
and E.~Frenkel [1,2] for description
of a space of local integrals of motion for
Toda field theories associated to affine Lie algebra ${\bf g}$ [3,4].
It was proved that the phase space of
the classical Toda theories was isomorphic
to the quotient $N_+/A_+$, where $N_+$ was the Lie group of a
nilpotent Lie subalgebra $\n_+$ of algebra ${\bf g}$, and $A_+$ is its
principle commutative subgroup.
This identification gave possibility to interprete the evolution of the Toda
system on its phase space in simple terms. Namely,
Toda Hamiltonian action on the phase space
coincided with the left infinitesimal action on the nilpotent subalgebra $\n_+$
on $N_+/A_+$ , and action of Toda local integrals of motion
 was given by the right infinitesimal action of a
commutative subalgebra $\a_-$, opposite to the principal
commutative subalgera $\a_+$ with respect to the Cartan involution.
The space of local integrals of motion of Toda theory
(quantum and classical) was ungerstood as a cohomology class of certain
complex.
We transfer this technique to the simpliest case of classical
Maxwell-Bloch (MB)
equation [3,5,6].

Here we will use the following simple form of classical MB equation
$$
\eqalign{
\d_\tau\beta&=\- ,
\cr
\d_\tau\gamma&=\+,}
$$
where $\beta(z),\,\,\gamma(z)$ are functions on the circle $|z|=1,\,\,
z\in\BC$.
MB eq. can be rewritten in the  zero-curvature form (compare to [5])
$$
\eqalign{
B_1&=\lambda\sigma_3+\left(\matrix{0&-q \cr r & 0}\right),
\cr
C_1&=-{1\over4}\lambda^{-1}\-
\left(\matrix{\gamma&-\gamma^2\cr 1&-\gamma}\right),
\cr
\d_\tau B_1&-\d C_1 - [B_1,C_1]=0 ,
}
$$
where $ \d = \d/\d z$ ,
$ \sigma_3 $ is Pauli matrix and $r, \,\,q $ are functions
$ r=\half\beta ,\,\,q =\half(-\beta\gamma^2+2\d\gamma)$.

MB eq. belongs to the AKNS hierarchy [3]
$$
B_n(\lambda)=\sum_{j=0}^n b_j \lambda^{n-j},\,\,\,
C_n(\lambda)= \sum_{j=0}^{n-1} c_j \lambda^{j-n},\,\,\,n=1,2,\cdots
$$
where $b_0=\sigma_3$ ; $b_j$ , $c_j$ are matrixes $2\times2$.
$$
\eqalign{
\d_{x_n} B_m &-\d_{x_m} B_n + [B_m , B_n] = 0,
\cr
\d_{y_n} C_m &-\d_{y_m} C_n + [C_m , C_n]=0,
\cr
\d_{y_n} B_m &-\d_{x_m} C_n + [B_m , C_n]=0,
}
$$
for $m, n= 1,2,\cdots$.
In MB eq. $\d_{x_1} \equiv \d ,\,\,\, \d_{y_1}\equiv\d_{\tau}$.
We call the first line of this system as positive
(with respect to the spectral parameter $\lambda$)
part of the AKNS hierarchy.

MB eq. is the Hamilton equation
$$
\d_\tau\beta=\{\beta,\,H\},\quad \d_\tau\gamma=\{\gamma,\,H\}
$$
with hamiltonian
$$
\oint d z \beta(z)\+ - \oint d z \gamma(z) \-
$$
and Poisson brackets
$$
\eqalign{
\left\{\gamma(x),\beta(y)\right\}
&=\delta(x-y)+\half\epsilon(x-y)\gamma(x)\beta(y),
\cr
\left\{\gamma(x),\gamma(y)\right\}
&=-\half\epsilon(x-y)\gamma(x)\gamma(y),
\cr
\left\{\beta(x),\beta(y)\right\}
&=-\half\epsilon(x-y)\beta(x)\beta(y),
}
$$
where $\epsilon$ is sign function.
In this paper we investigate local integrals of motion of MB eq., i.e.
local functionals of the form
$
\int d z P(\beta(z), \gamma(z)),
$
where $P$ is a differential polinomial in $\beta,\,\gamma$.
It is known that local integrals of motion of MB eq. coincide
with the hamiltonians of the positive part of the AKNS hierarchy.

Consider affine algebra $\sl$ without central extension as defined over
formal Laurent power series
{\bf C}$((t))$: $\g=sl_2\otimes{\bf C}((t))\oplus\BC d$,  $d$ is standart
{\bf Z}-grading operator, $d = t d/dt.
 $
Let $e, h, f$ be standart $sl_2$ generators; then define Cartan subalgebra
$\h$
of the affine algebra $\g$:
$\h=\BC h\otimes 1\oplus\BC d$. Nilpotent subalgebras
$\n_+(\n_-)$ of $\g$ are generated by $e,f\otimes t$
(resp. $f,e\otimes t^{-1}$).
For $\g$ we have Cartan decomposition
$
\g=\n_+\oplus \h \oplus \n_-.
$
Denote $\h_\pm$ as commuting subalgebra of $\n_\pm$, generated by
$h_{\pm i}=h\otimes t^{\pm i},\quad i=1,2,\cdots$.
Let $N_+$ be the Lie group of the nilpotent subalgebra $\n_+$ and
$H_+$ be the Lie group of the commutative subalgebra $\h_+$.

We will show that the MB phase space is isomorphic to the quotient
$N_+/H_+$.

The action of two parts of MB hamiltonian on the phase space is
identified with the left infinitesimal
action of the nilpotent subalgebra $\n_+$
on the quotient $N_+/H_+$. Then we interpret the space
of local integrals of motion of MB eq. as the
first cohomology of $\n_+$ with coefficients in $\BC(N_+/H_+)$.
This formulation allows to describe the space
of the MB local integrals
of motion and gives
 opportunity to quantize them [1].
Moreover the action of local integrals of motion
(i.e. positive AKNS hamiltonians)
on the phase space can
be realized as the right infinitesimal action of the
commutative subalgebra $\h_-$
on the quotient $N_+/H_+$.

This paper is organized as follows. In Section 2 we introduce the MB phase
space as reduced  bosonic $\beta, \gamma, \phi$ system.
The MB hamiltonian  is written. We show that an action of
the hamiltonian gives rise
to an action of nilpotent subalgebra $\n_+$ of $\g$ on the phase space.
In Section 3 geometrical picture of MB eq. is given.
We prove that there exists the $\n_+$-isomorphism between
MB phase space and quoutient $N_+/H_+$. In Section 4 we give cohomological
description of the space of local integrals of motion. In Section 5
we show that vector fields generated by the local integrals of motion of MB eq.
are given
by the right infinitesimal action of $\h_-$ on $N_+/H_+$.
\medskip\noindent
{\bf 2. Definition and Hamiltonian Picture}
\medskip\noindent
In this section we give the MB eq. as a reduction of the
$\beta, \gamma, \phi$ system.
Consider three classical bosonic fields $\beta(z),\gamma(z),
\d\phi(z)$ on circle $z\in {\bf C}$ , $|z|=1$ with  Poisson brackets
$$
\eqalign{
\left\{\gamma(z),\beta(w)\right\}
&=\delta(z-w),
\cr
\left\{\phi(z),\phi(w)\right\}
&=\half\epsilon(z-w).
}\eqno(1)
$$
Let $\Pi_0$ be space of differential polinomials in $ \beta,\gamma,\d\phi$,
i.e. polinomials of the form
$$
P(\beta(z),\gamma(z),\d\phi(z),\d\beta(z),\d\gamma(z),\d^2\phi(z),\cdots),
$$
$(\d=\d/\d z)$
We will refer to
$$  F[\beta,\gamma,\phi]=\oint dz\,P\left(\beta(z),\gamma(z),
\d\phi(z),\d\beta(z),\d\gamma(z),\d^2\phi(z),\cdots\right),$$
as local functionals. Denote space of local functionals as $\widehat\Pi_0$.
The space  $\widehat\Pi_0$ can be identified with
$\Pi_0/\d\Pi_0$ , because integral of total derivative is zero.
We can ever consider space of functions of the form
$P(\beta,\gamma,\d\phi,\cdots)e^{n\phi}$, where $P\in\Pi_0$,
which can be treated as $\Pi_n=\Pi_0\otimes e^{n \phi}$.
Denote $\d'=\d+ n \d\phi$. It is obvious that
$\d ( P e^{n \phi}) = (\d'P) e^{n \phi}$ , so the action of $\d$ on $\Pi_n$
is $(\d + n\d\phi)\otimes 1$.
Define $\widehat\Pi_n$ as integrals of elements from $\Pi_n$ , i.e.
$\widehat\Pi_n = \Pi_n/\d\Pi_n$.

Introduce a hamiltonian
$$H=\oint dz\,\beta(z)e^{\phi(z)}\,-\,\oint dz\,\gamma(z)e^{-\phi(z)}\in
\widehat\Pi_1\oplus\widehat\Pi_{-1}
\eqno(2)$$
Consider the Hamilton equation with  Poisson brackets (1) and
hamiltonian (2)
$$\eqalign{
\d_\tau\beta&=\{\beta,\,H\}=  e^{-\phi},\cr
\d_\tau\gamma&=\{\gamma,\,H\}=e^{\phi},\cr
\d_\tau\d\phi&=\{\d\phi,\,H\}=\beta e^\phi+\gamma e^{-\phi},}
\eqno(3)
$$
where $\d_\tau$ designates the derivative over the time $\tau$.

Define the action of operators $\bar Q_1 , \bar Q_0  $ on the space $\Pi_0$
$$
\eqalign{
\bar Q_1 : \,\Pi_0\longrightarrow \Pi_1\,\,,\,\,\,  \bar Q_1 &=
\left\{\cdots,\oint \, dz\,\beta e^\phi \right\} ,
\cr
\bar Q_0 :\,\Pi_0\longrightarrow \Pi_{-1}\,\,,\,\,\,   \bar Q_0 &=
\left\{\cdots,\oint\,dz\,\gamma e^{-\phi}\right\} .
}
\eqno(4)
$$
It is easy to see that operators $\bar Q_1 , \bar Q_0 $  commute with
the action of $\d$. So the action of these operators is well- defined on
$\widehat\Pi_0$.

Comparing definition (4) with hamiltonian (2) , we see that
evolution (3) is given by the operator
$$
\d_\tau=\bar Q_1 -\bar Q_0
\eqno(5)
$$

A local functional $I\in\widehat\Pi_0$ is called a local integral of motion
of the system (3), if $\d_\tau I = 0$.The local integrals of motion
will be designated as IM.
We see from definition (4) that the space of
IM of system (3) can be written as:
$$
\widehat\Pi_0\supset \hbox{the space of IM of system (3)}
=\hbox{ Ker }\bar Q_1 \cap\hbox{ Ker }\bar Q_0 ,
$$
because the operator $\d_\tau$ maps an element from $\widehat\Pi_0$ to
an element from the sum of two different spaces $\widehat\Pi_1\oplus
\widehat\Pi_{-1}$.

Introducing translator operator T
$$
T:\,\,\Pi_n\longrightarrow\Pi_{n+1}\, ,\,\,
P\otimes e^{n\phi}\longrightarrow P\otimes e^{(n+1)\phi},
$$
where $P\in\Pi_0$,
define vector fields $ Q_1,\,Q_0 $ on $\Pi_0$
$$
\eqalign{
Q_1 &=\bar Q_1 T^{-1}
\cr
Q_0 &=\bar Q_0 T
}
\eqno(6)
$$

Denote the expression $\beta\gamma-\d\phi$ as $\j$.
The evolution (3) of the function $\j$ is trivial
$$
\d_\tau\j(z) = 0
$$
Thus we can reduce the space $\Pi_0$, imposing the condition
$$
\j(z) = 0 ,
\eqno(7)
$$
which is compatible with evolution (3).
Reduction (7) of the system (3) gives Maxwell-Bloch equation
$$
\eqalign{
\d_\tau\beta&=\- ,
\cr
\d_\tau\gamma&=\+.}
\eqno(8)
$$

Now we describe Poisson structure of  $ \beta, \gamma, \d\phi $
system reduced by
the condition $\j=0$. Let ue define
$$
\pi_0=\Pi_0/\hbox{ D Pol}(\j) \Pi_0
$$
where D Pol is a set of differential polinomials.

Namely, a differential polinomial
$P(\beta,\gamma,\d\phi )$ from the space $\pi_0$ is
equivalent to
$$
P(\beta, \gamma,\d\phi)\sim P(\beta, \gamma, \d\phi )+
\sum_{n=0}^\infty R_n\,\d^n\j,
\eqno (9)
$$
where $R_n$ is differential polinomial in $\beta, \gamma, \d\phi$.
Introduce a Dirac bracket on $\pi_0$ $\left\{\,,\,\right\}^*$
with the main property
$$
\left\{P(\beta,\gamma,\d\phi),\j\right\}^*=0,
$$
for any differential polinomial P. Then we have:
if A, B are differential polinomials in $ \beta,\gamma,\d\phi$, such that
$A\sim B $ , then $ \left\{\xi,A\right\}^*\sim\left\{\xi,B\right\}^*$
for arbitrary $\xi$.
Using this property and equivalence formula (9)
we can treat $\pi_0$ as Poisson manifold with coordinates
$\beta,\gamma,\d\beta,\d\gamma$ etc and Poisson-Dirac bracket:
$$
\eqalign{
\left\{\gamma(x),\beta(y)\right\}^*
&=\delta(x-y)+\half\epsilon(x-y)\gamma(x)\beta(y),
\cr
\left\{\gamma(x),\gamma(y)\right\}^*
&=-\half\epsilon(x-y)\gamma(x)\gamma(y),
\cr
\left\{\beta(x),\beta(y)\right\}^*
&=-\half\epsilon(x-y)\beta(x)\beta(y),
}
\eqno (10)
$$
where $\epsilon$ is sign function.

Poisson structure (10) is the first Poisson structure $\{\,\,,\,\,\}_1$
for NLS eq. [7].
After reduction (7) space $\Pi_n$ transforms to $\pi_n$
$$\pi_n=\pi_0\otimes\hbox {exp}(n\int^z d \tau\, \beta\gamma).
$$
The action of derivative $\d$ on $\pi_n$ is following
$$
(\d+n\beta\gamma)\otimes 1
\eqno(11)
$$

Define $ \hat {\pi}_n=\pi_n/\d\pi_n$ as a space of functionals of the form
$$
\oint dz\,\hbox{ D Pol} (\beta,\gamma)e^{n\int^z dt\,\beta\gamma}.
$$
After reduction (7) hamiltonian (2) transforms as follows
$$
\oint dz\,\beta(z)\+\,-\,\oint dz\,\gamma(z)\-\,\,
\in \hat \pi_1\otimes\hat\pi_{-1}.
\eqno(12)
$$

The action of operators $\bar  Q_1   $ and $\bar Q_0  $ is compatible with
reduction (7) , i.e. $\bar Q_1 \j=\bar Q_0 \j=0$. Thus they can be limited
from $\Pi_0$ to $\pi_0$
$$
\eqalign{
\bar Q_1 : \,\pi_0\longrightarrow \pi_1\,\,,\,\,\,
\bar Q_1 &=\left\{\cdots,\oint dz\,\beta\+\right\}^*,
\cr
\bar Q_0 : \,\pi_0\longrightarrow \pi_{-1}\,\,,\,\,\,
\bar Q_0 &=\left\{\cdots,\oint dz\,\gamma\-\right\}^*.
}
\eqno (13)
$$
The explicit formula for the action (13) of operators
$\bar Q_1$ and $\bar Q_0$  on the element $P\in\pi_0$ is
$$
\eqalign{
\bar Q_1 P&=\sum_{n\geq0} B_n^+ {\d P\over\d(\d^n\gamma)}\otimes \+,
\cr
\bar Q_0 P&=\sum_{n\geq0} B_n^- {\d P\over\d(\d^n\beta)}\otimes\-,
}
\eqno(14)
$$
where $ P,B_n^\pm \in \pi_0$ and
$$
\eqalign{
\d_z^n\+&=B_n^+\+,
\cr
\d_z^n\-&=B_n^-\- .
}
$$

The MB eq. (8) is treated as Hamilton equation
$$
\eqalign{
\d_\tau\beta&=\{\beta,\,H\}^* ,
\cr
\d_\tau\gamma&=\{\gamma,\,H\}^*
}
$$
with brackets
$\left\{\,\,,\,\,\right\}^*$ (10) and hamiltonian (12).

A local functional $I\in\widehat\pi_0$ is called a local integral of motion
of MB eq., if $\d_\tau I = 0$. After imposing the reduction (7)
on the formula for evolution (5)
we get that
the space of IM of MB eq. *(8) is the intersection of kernels of operators
$\bar Q_1$, $\bar Q_0$
$$
\hat\pi_0\supset \hbox{ the space of IM of MB eq.}
=\hbox{ Ker } \bar Q_1\cap \hbox{ Ker }\bar Q_0
$$
Using translator operator T
$$
T:\,\,\pi_n\longrightarrow\pi_{n+1}\, ,\,\,
P\otimes e^{n\int^z dt\,\beta\gamma}
\longrightarrow P\otimes e^{(n+1)\int^z dt\,\beta\gamma},
$$
where $P\in\pi_0$, and explicit formula (14)
define vector fields $ Q_1,\,Q_0 $ on $\pi_0$
$$
\eqalign{
Q_1 &=\bar Q_1 T^{-1}=\sum_{n\geq0} B_n^+{\d\over\d(\d^n\gamma)},
\cr
Q_0 &=\bar Q_0 T=\sum_{n\geq0} B_n^-{\d\over\d(\d^n\beta)}.
}
\eqno(15)
$$

Vector fields $Q_1 $ and $Q_0$ on $\pi_0$ satisfy Serre relations
for  the nilpotent subalgebra $\n_+$
$$
ad_{Q_0}^3\cdot Q_1=0\,\, \hbox{ and }\,\, ad_{Q_1}^3\cdot Q_0=0
$$
and can be identified with the generators
of the nilpotent subalgebra $\n_+$ : $e$ and $f\otimes t$.

Thus vector fields $Q_1 $ and $Q_0$ give structure of $\sl$ module to $\pi_0$.

Define grading on $\pi_0$:
$$
\eqalign{
deg\beta&=(s=-1,q=1),\,\,\, deg\gamma=(s=0,q=-1),
\cr
deg\d&=(s=-1,q=0),\,\,\, deg v_n=(s=-\half n (n-1),q=n),
}
\eqno (16)
$$
where q is the isospin with respect to current $\j$:
$ \left\{\j(x),P_q(y)\right\}=q \delta(x-y) P_q(y)$

Then $deg \bar Q_1^*=deg \bar Q_0^*=0$.
\medskip\noindent
{\bf 3. Geometrical Picture}
\medskip\noindent
In this section we give geometrical description of the space $\pi_0$.
Recall that we consider affine algebra $\sl$ without central extension
as defined over
formal Laurent power series
{\bf C}$((t))$: $\g=sl_2\otimes{\bf C}((t))\otimes d$.
For $\g$ we have Cartan decomposition
$
\g=\n_+\oplus \h \oplus \n_-,
$
where $\n_+$ is the nilpotent subalgebra
$\n_+=\BC e\otimes 1\oplus sl_2\otimes \BC[[t]]$,
$\n_-$ is the opposite nilpotent subalgebra
$\n_-=\BC f\otimes 1\oplus sl_2\otimes \BC[t^{-1}]$.
Let $G$ be the Lie group of the affine algebra $\g$,
$N_\pm$ be the Lie group of the nilpotent subalgebra $\n_\pm$,
$B_\pm$ be the Lie group of the Borel subalgebra ${\bf b}_\pm=\n_\pm\oplus\h$.

The group $N_+$ is isomorphic to the big cell X of the flag manifold
$F=B_-\backslash G$, which is the orbit of 1 under the action of $N_+$.
See references [8-10].
The Lie algebra $\g$ acts infinitesimally  from the
right by vector fields on $F$ and hence on $N_+$. So does the Lie algebra
$vect_-=\BC[t^{-1}] t \d_t$, with generators $L_n=t^{-n+1} \d_t,\,\,n\geq 0$.
Denote by $\nu$ the Lie algebra of vector fields on $N_+$. It contains two
commuting Lie subalgebras: $\n_+^R$ and $\n_+^L$ of vector fields of the right
and the left infinitesimal action of $\n_+$ on its Lie group.
The vector field of the left
infinitesimal action of $\beta\in \n_+ $ on $N_+$ denoted by $\beta^L$.
The Lie algebra $\n_+^R$ is a part of a larger subalgebra of $\nu$, which is
isomorphic to $\tilde{\bf g}=\g\times vect_-$. The vector field of the right
infinitesimal action of $\alpha \in\tilde{\bf g} $ on $N_+$ will be denoted by
$\alpha^R$.

For $\l\in \BC$ let $M_\l$ be Verma module over $\g$ of the
$sl_2$ spin $\l$ with the highest vector $v_j$
$$
\n_+\cdot v_j=0,\quad h\cdot v_j=j v_j,\quad M_\l=U(\n_-)\cdot v_j.
$$

$\ms$ is the module contragradient to $\m$ with pairing  $<\,,\,>:
\ms\times\m\rightarrow\BC$. Following [2] we describe a geometrical
construction of $\ms$.
Let $\omega$ be Cartan anti-involution on $\g$,
mapping $e, f\otimes t$ to $f, e\otimes t^{-1}$ .

Define the right action of  $y\in \g$ on $x\in \ms$ as follows:
$$
<x\cdot y, z>=<x,\omega(y)\cdot z>, \quad z\in\m
$$

The module $\ms$ can be identified with space of functions $\BC(X)$
on the big cell X with respect to a
twisted action. The right action of $\beta\in \g$
on $\ms$ gives under this identification the action of
$$
\beta^R + j F(\beta)\,\, \hbox{ on }\,\, \BC(X),
$$
where $F(\beta)$ is function on X. We have
$F(h)=1$ and $\f=0$ for
 $\beta\in \n_+$.

Let vector $v_m$ be singular vector of Verma module $\m$ ,
$v_m=P\cdot v_\l$ for some element $P\in U(\n_-)$ and
$R\cdot v_m=0$ for any element $R\in U(\n_+)$.
This singular
vector defines homomorphism of $\g$-modules
$$
i_P: M_m\rightarrow M_\l,\quad u\cdot v_m\rightarrow(uP)\cdot v_\l
\eqno(17)
$$
for any $u\in U(\n_-)$. The map $i_P$ commutes with the $\g$-action and
is called intertwining opertor.

The left action of the element $\beta\in \n_+$ on $x\in \ms$ can be defined
as follows
$$
<\beta\cdot x, u\cdot v_\l>=<x,(u \omega(\beta))\cdot v_\l>,
\,\,\, u\in U(\n_-)
$$

Let $\bar P$ be the image of $P\in U(\n_-)$ under isomorphism
$U(\n_-)\rightarrow U(\n_+)$, which maps generators $e, f\otimes t$ to
$f, e\otimes t^{-1}$.
The homomorphism $\n_+\rightarrow \nu$, mapping $\alpha\in \n_+$
to $\alpha^L$ , can be extended uniquly to homomorphism from
$U(\n_+)$ to the algebra of differential operators on X . Let
$u^L$ be image of $u\in U(\n_+)$.

It is known [2] that homomorphism $i_P^*:\,\,\ms\rightarrow M_m^*$
dual to (17) can be realised as differential operator $\bar P^L$ on X.
For example, consider the map $i_f: M_{-2}\rightarrow M_{0}$,
 $ u\cdot v_{-2}\rightarrow(u f)\cdot v_{0}$. Then dual map
$i_f^*: M_{0}^*\rightarrow M_{-2}^*$ is given by the left
infinitesimal action $e^L$ on $\BC(X)$ treated as $M_0^*$.

Let us study further the left action of $U(\n_-)$ on $\BC(X)$.
In order to simplify formulas introduce
the Chevalle basis of $\n_+$ :  $e_1=e, e_0=f\otimes t$.
For $\f ,\,\,\beta\in \sl $ we have following
$$
[e_i^L,\beta^R]=2 (-)^i F(\beta) e_i^L ,\quad i=0,1 .
\eqno (18)
$$
For proving (18) it's sufficient to consider $e_1^L$ as $\g$-homomorphism
from $M_0^*$ to $M_{-2}$ and $e_0^L$ as $\g$-homomorphism
from $M_0^*$ to $M_{2}$ .
Here we treat $\ms$ as $\BC(X)$ with twisted
action.
After identifing $\BC(N_+)$ with $(U(\n_+))^*$ introduce grading on
$\BC(X)\simeq\BC(N_+)$ with respect to degree of $t$ and action of $\h$
$$
deg\,e\otimes t^n=(n,1),\,\,deg\,f\otimes t^n=(n,-1),\,\,
deg\,h\otimes t^n=(n,0)
\eqno (19)
$$

We prove that the space of functions $\BC (\Y)$
on the homogeneous space $\Y$
is isomorphic to $\pi_0$ as $\n_+$-modules.
Indeed, vector fields $Q_1,\,\,Q_0$ (15) define structure of $\n_+$-module
on $\pi_0$. Let $x_i,\,\,y_i$ be coordinates on $\pi_0$
$$
x_{i+1}=\d^i \beta,\quad y_i=\d^i \gamma,\quad i\geq 0
$$
Then for the vector fields $Q_1,\,\,Q_0$ on $\pi_0$ we have (see (15))
$$
\eqalign{
Q_1&={\d\over {\d y_0}}+ x_1 y_0 {\d\over {\d y_1}}+
(x_1^2 y_0^2+x_2 y_0+x_1 y_1){\d\over {\d y_2}}+\cdots
\cr
Q_0&={\d\over {\d x_1}}-x_1 y_0{\d\over {\d x_2}}  +
(x_1^2 y_0^2-x_2 y_0+x_1 y_1){\d\over {\d x_3}}+\cdots
}
\eqno (20)
$$

Let X be an operator on $\pi _0$ of the form $\sum_i (X_i \d/\d x_i+
Y_i \d/\d y_i)$ , then define its shift term as all the terms
$X_i \d/\d x_i$ or $Y_i\d/\d y_i$, for which $X_i$ or $Y_i$ is constant.
{}From (20) shift terms of $Q_1$ and $Q_0$ are $\d/\d y_0$ and
$\d/\d x_1$. The shift term of $[Q_1,Q_0] $ is 0.  $Q_1$ and
$Q_0$ satisfy Serre relation and generate $\n_+$. In our notation
we can treat $Q_1$ as $e$ and $Q_0$ as $f\otimes t$. It's easy to see
that vector fields corresponding to subalgebra $\h_+$ have no shift terms,
and shift term of $e\otimes t^n$ is $\d/\d y_n,\quad n\geq 0$, while
shift term of $f\otimes t^n$ is $\d/\d x_n,\quad n\geq 1$.

Consider the module $\pi_0^*$ over $\n_+$, dual to $\pi_0$. We can identify
$\pi_0$ with $\pi_0^*$ as linear spaces, choosing the monomials
$x_{k_1}/k_1 !\cdots x_{k_n}/k_n !$ and $y_{k_1}/k_1 !\cdots y_{k_n}/k_n !$
as an ortonormal basis. The formulas for the action of $\n_+$ on $\pi_0^*$
are obtained from the formulas for its action on $\pi_0$ by interchainging
$x_n\,\,(y_n)$ and $\d/\d x_n$ (resp.$\d/\d y_n$). Combinations of
$Q_0,\,\,Q_1$ , corresponding to $\h_+$ , act on $1^*\in \pi_0^*$ by 0
(because they have not shift terms).

Let us introduce
$$
N=U(\n_+)\otimes_{U(\h_+)} \BC,
$$
$\n_+$-module, induced from the trivial one-dimentional representation
of subalgebra $\h_+$.
Since action of $\h_+$ on $1\otimes 1 \in N$ is trivial, there is unique
$\n_+$-homomorphism $N\rightarrow \pi_0$, sending $1\otimes 1\in N$ to
$1^*\in \pi_0^*$ , and
$(e\otimes t^{n_1})\cdots( e\otimes t^{n_k})( f\otimes t^{m_1}) \cdots
( f\otimes t^{m_l})\otimes 1$ maps to
$y_{n_1}\cdots y_{n_k} x_{m_1}\cdots x_{m_k} \cdot 1^*$ + lower order terms.
Therefore map $N\rightarrow\pi_0$ has no kernel. $N$ and $\pi_0$ coinside
as linear spaces with respect to graiding (16) and (19).
Thus
$$
\pi_0 \simeq\left(U(\n_+)\otimes_{U(\h_+)}\BC\right)^*\simeq\BC(N_+/H_+)
\eqno(21)
$$
as $\n_+$ -modules.

Recall that $\h_\pm$ is commuting subalgebra of $\n_\pm$, generated by
$h_{\pm i}=h\otimes t^{\pm i},\quad i=1,2,\cdots$.
Consider eq. (18) for $\beta=h_{-1}$
$$
[e_i^L,h_{-1}^R]=(-)^i\, 2 F(h_{-1})e_i^L, \quad i=0,1
\eqno (22)
$$
as vector fields on $\BC(N_+)$.
For the function $F (h_{-1})$ on the group $N_+$ we have
$$
x_+^R\cdot F(h_{-1})=0,
\eqno (23)
$$
for $x_+\in \h_+$.
This can be derived from commutator (22). Indeed,
$$
[x_+^R,[e_i^L,h_{-1}]]=-[e_i^L,[h_1^R,x_+^R]]-[h_{-1}^R,[x_+^R,e_i^L]]=0.
$$
So for $i=0, 1$  $(x_+^R\cdot F(h_{-1}))\cdot e_i^L=0$ and then (23).
Eq. (23) gives us example of $\h_+^R$- invariant functions on $\BC(X)$.
It's obvious that the right action of $\h_-$ on $F(h_{-1})$ preserves
$\h_+$-invariancy, thus $\beta^R\cdot F(h_{-1})$ for $\beta\in \h_-$ is
$\h_+$ -invariant. Other families of $\h_+$ invariant functions on $\BC(X)$
are given by $F(f)$ , $F(e\otimes t^{-1})$ and the right action of $\h_-$
on them.

Thus we can treat (22) as vector fields
on $\coset$, because $e_i^L, h_{-1}^R$ commutes with $\h_+^R$, and $F(h_{-1})$
is $\h_+$-invariant.

Using isomorphism (21) we can identify $e_i^L$ acting on
$\coset$ with $Q_i$ acting on $\pi_0$. The element
$F(h_{-1})$ of $\coset$
has grading (16): $deg F(h_{-1})=(-1;0)$. The only image of $F(h_{-1})$
in $\pi_0$ under isomorphism $\BC(\Y)\rightarrow\pi_0$ of the same degree
is $\,\,{const}\cdot x_1 y_0(=\,\,{const}\cdot\beta\gamma)$ ,
$deg x_1 y_0=(-1;0)$.

So using (22) and isomorphism (21), we have
$$
[Q_i, \eta_{-1}]=\,\,{const}\cdot(-)^i\,2 x_1 y_0 Q_i , \quad i=0,1
\eqno (24)
$$
as vector fields on $\pi_0$, where $\eta_{-1}$ is unknown vector field
on $\pi_0$.

We prove that the vector field $\eta_{-1}$ is proportional to the vector
field of derivative $\d=\sum x_i\d/\d x_{i-1}+y_i\d/\d y_{i-1}$.
If we choose $\,\,const\,=-\half$ then
$$
\eta_{-1} = \d
\eqno(25)
$$
on $\pi_0$.
Indeed, operators $\bar Q_1 =T Q_1$ and $\bar Q_0 =T^{-1} Q_0$
commute the action of $\d$ (11) on $\pi_0$.
That is why
$$
[Q_1,\,\d]=T^{-1}\,[\d,\,T]\,Q_1=x_0 y_1 Q_1 \,\hbox{ and }\,\,
[Q_0,\,\d]=T\,[\d,\,T^{-1}]=-x_0 y_1 Q_0
$$
So the image of the action of $h_{-1}^R$ on $\BC(\Y)$ under isomorphism (21)
is the action of $\d$ on $\pi_0$.
\medskip\noindent
{\bf 4. Cohomology Computation}
\medskip\noindent
In this section we describe the space of IM of MB eq. using
a double complex [2].
Let $\b$ be the dual of BGG resolution for $\sl$ [11]
$$
\b=\bigoplus_{j\geq O}\bj,
$$
where $B^0(\sl)=M_0^*$ and  $\bj=M_{2j}^*\oplus M_{-2j}^*$.
Here $M_j^*$ is contragradient
module to the Verma one with $sl_2$ spin $j$ and level k=0.

The Verma module $M_0$ contains singular vectors, labeled by the
elements of the Weyl group of $\sl$. Denote this vectors by $w_0$ and
$w_{\pm j}$, $j=1,2,\cdots$. The weights of $w_{\pm j}$ are $\pm 2j$.
The action of $U(\n_-)$ on $w_{\pm j}$ generates the submodule
$M_{w_{\pm j}}$ isomorphic to $M_{\pm 2j}$. Let $l(w_{\pm j})=j$.
Choose two elements $w\,,\,w'$ of the Weyl group, such that
$l(w)=l(w')+1$. Then $\bar P_{w,w'}^L$ is a map, dual to embeding
(17) $i_{w,w'}$: $M_w\rightarrow M_{w'}$. (We identify $\ms$ with
$\BC(X)$).

The 0-th cohomology of $\b$ is one-dimentional and all higher ones vanish.
Differentials $\delta^j: B^j(\sl)\rightarrow B^{j+1}(\sl)$ of $\b$
can be written in a common way:
$$
\delta^j=\sum_{l(w)=j, l(w')=j+1} \epsilon_{w,w'} \bar P_{w,w'}^L.
$$
{}From this formula the first differential $\delta^1$, mapping
$M_0^*\rightarrow M_2^*\oplus M_{-2}^*$ is $e_1^L - e_0^L$
as the action on $\BC(X)$.

The right action of $\sl$ on complex $\b$ commutes with the differentials
$\delta^j$. This valid for $\h_+\subset \sl$. So we can take quotient of $\b$
by the right action of Lie subgroup $H_+$ of $N_+$. Denote this complex by
$\F$:
$$
\F=\bigoplus_{j\geq 0} F^j(\sl),
$$
where $F^0(\sl)=\pi^{(0)}$ and
$F^j(\sl)=\pi^{(2j)}\oplus\pi^{(-2j)}$, here $\pi^{(\pm 2j)}$ denotes
space of $\h_+$-invariants of $M^*_{\pm 2j}$.

The right action of $\h_-$ on $\b$ gives rise to $\h_-$-action on $\F$,
because for $x\in \h_-$ function $F(x)$ is $\h_+$-invariant .
The action of $ x\in \h_-$ on $\pi^{(j)}$ is given by
$$
x^R + j F(x)
\eqno (26)
$$
This action commutes with the differentials of the complex $\F$, that is why
the action of $\h_-$ is defined on the cohomologies of $\F$.

For description of the space
of IM of MB eq.  cohomologies of $\F$ should be computed.
We prove that
cohomologies of $\F$ are isomorphic to
$\land^*(\h_+^*)$.

Since $\b$ is injective resolution of the trivial representation of
$\n_+$ , the cohomologies of $\F$ coincide with
the cohomologies $H^*(\n_+,\pi_0)$ of the Lie algebra $\n_+$ with coefficients
in the module $\pi_0$
[1].
Because $\pi_0$ can be identified with
$\left(U(\n_+)\otimes_{U(\h_+)}\BC\right)^*$, we have
by Shapiro lemma:
$$
H^*(\n_+,\pi_0)\simeq H^*(\h_+,\BC)\simeq\land^*(\h_+^*)
\eqno (27)
$$

In [2] it was proved that the action of $\h_-$ on $H^*(\n_+,\pi_0)$ is trivial.
In particular, the operator $h_{-1}$ acts trivially on the cohomologies.
We already know that its action on $\pi_0$ consides with the action of
$\d $. Consider now the action of $h_{-1}$ on $\pi^{(\pm 2j)}$.
It is given by $h_{-1}^R \pm 2 j F(h_{-1})$. The image of $F(h_{-1})$
under isomorphism $\BC(\Y)\rightarrow\pi_0$ is equal to $-\half\beta \gamma$.
So the action of $h_{-1}$ on $\pi^{(\pm 2j)}$
coincides with the action (11) of derivative $\d$ on $\pi_{\mp j}$  and
we have isomorphism $\pi^{(\pm 2j)}=\pi_{\mp j}$ with respect to the
action of $\n_+$.
Thus $F^0(\sl)=\pi_0$ , $F^j(\sl)=\pi_{-j}\oplus\pi_{j}$ and
the first differential $\delta^1$: $\pi_0\rightarrow\pi_{-1}\oplus\pi_{1}$
is equal to $\delta^1=\bar Q_1- \bar Q_0$. We see that differential
$\delta^1$ coincides with the action of MB hamiltinian on $\pi_0$.
This observation enables us to compute the space of IM of MB eq.
Indeed, the space of IM is isomorphic to the kernel of the map
$\delta^1=\bar Q_1^*- \bar Q_0^*$:
$\hat \pi_0\rightarrow\hat\pi_{-1}\oplus\hat\pi_{1}$.
Recall that $ \hat {\pi}_n=\pi_n/\d\pi_n$.
In order to use the complex $\F$ for computing IM we should get rid of
total derivatives in $z$, i.e. consider $\F/\d\F$. For these reasons
a double complex is used.

Now the main result can be formulated : the space of integrals of motion of the
MB eq. (8)
is lineary spanned by elements $H_m$, $m=1,2,\cdots$, where
$degH_m=(-m,0)$.

 Since $\d=h_{-1}$ commutes with the differentials of $\F$, we can
consider the doube complex:
$$
\BC\longrightarrow\F \mathop{\longrightarrow}\limits^{\pm h_{-1}}\F
\longrightarrow\BC
\eqno(28)
$$

Using the spectral sequence, in which $h_{-1}$ is the 0 th
differential , one gets that
the 1st cohomology of the double complex $H^1_{tot}$ is isomorphic
to the space of IM. We can also compute this cohomology using
the other spectral sequence. Because  $h_{-1}$ acts
trivially on $H^1(\F)$, $ H^1_{tot}\simeq H^1(\F)\simeq \h_+^*$.
Therefore space of IM is linearly spanned by elements $H_m$ of
degree (s=-m , q=0) for m=1,2$\cdots$.
\medskip\noindent
{\bf 6. Construction of IM and vector fields}
\medskip\noindent
Now we show explicitly how IM are connected with the first cohomology
class $H^1(\F)$.
Consider the element ${\cal H}$ of cohomology class
$H^1(\F)\subset \pi_{-1}\oplus\pi_1$. If we apply $\d$
to ${\cal H}$, we obtain a trivial cycle. So there exists element $h\in\pi_0$
such that
$$
\delta^1\cdot h=\d{\cal  H},
\eqno(29)
$$
where $\delta^1$ is the first differential of the complex $\F$ equal to the
MB flow:
$\delta^1=\bar Q_1 - \bar Q_0 = \d_\tau$.
By construction $h$ is not total derivative, so it belongs to IM.
Thus, formula (29) explicitly gives isomorphism between the spase of IM
and the cohomology class $H^1(\F)$.

For example, consider two lowest IM:
$h^{(-2)}=r q\in \pi_0$ and $h^{(-3)}=r\d q\in\pi_0 $
of degrees (-2; 0) and (-3; 0).
They are connected via formula (29) with
$j^{(-1)}={1\over4} \gamma\-\in H^1(\F) $
and $j^{(-2)}={1\over 4} \beta\gamma^2\-\in H^1(\F)$
of degrees (-1; 0) and (-2; 0).

Denote by $H_m\in\hat\pi_0$ element of the space of IM of MB eq. of degree
(-m; 0). Let $\eta_m$ be the vector field on $\pi_0$:
$$
\eta_m=\left\{\cdots,H_m \right\}^*.
\eqno(30)
$$
It is easy to see that $\eta_1=\d$.
We can treat $\eta_m$ as vector fields on $\Y$ under isomorphism
$\pi_0\simeq\BC(\Y)$.  On the other hand the right
action of generator $h_{-m} ,\quad m=1,2,\cdots$ of the subalgebra
$\h_-\in\sl$ on $\Y$ also
defines vector fields $\mu_m$ on $\Y$.
Now we formulate the result : the vector fields $\mu_m$ coincide with $\eta_m$,
$m=1,2\cdots$.

The statement was already proved for m=1.
We have shown that action of $h_{-1}$ on $\pi_0$ gives us $\d$.
So does vector field $\eta_1=\{\cdots,H_1\}^*$, where
$H_1=\oint d z r q$.
Consider now commutator of operators $\bar Q_1$ and $\eta_m$,
acting on $P\in \pi_0$
$$
[\bar Q_1,\eta_m] P =\left\{\{\oint d z \beta\+, H_m\}^*, P\right\}^*=0,
$$
since $\{\oint d z \beta\+, H_m\}^*=0$ by the definition of IM.
The same is valid for $\bar Q_0$.

Now we can compute commutator of vector fields $ Q_i$ and $\eta_m$
on $\pi_0$. Using expressions for
$\bar Q_1 = T Q_1$ and $\bar Q_0 = T^{-1} Q_0$
one can derive:
$$
[Q_i,\eta_m]=-f^i_m Q_i,
\eqno(31)
$$
as vector fields on $\pi_0$, where $f^i_m$ ---certain function on $\pi_0$.

Let us consider now eq.(18) on $\BC (N_+)$ and vector fields $\beta$ on
$\BC(N_+)$ , satisfing it. It is obvious that, if vector fields $\alpha$
and $\beta$ satisfy (18), then $[\alpha, \beta]$ satisfies it too.
So we have Lie algebra of vector fields on $\BC(N_+)$ , satisfing (18).
In [2] it was shown, that this Lie algebra is isomorphic to $\tilde {\bf g}$.
The only vector fields, satisfing (18), which in addition $\h_+$-invariant,
are generated by right action of $\h_-$ on $\BC(N_+)$, i.e. $\mu_m$.
On the other hand vector fields $\eta_m$ can be lifted to $\h_+$-invariant
vector fields $\tilde\eta_m$ on $N_+$, which satisfy relation (31) and thus
(18).
Comparing
the degrees of vector fields $\tilde\eta_m$ and $\mu_m$ we get the statement.

Now  proof of well- known result [7]: flows generated by IM of MB eq.
(i.e. the positive AKNS hamiltonians)
commutes with each other
$\{H_m,H_n\}=0$ for $ m,n=1,2,\cdots$
is obvious.

 Because of commutativity of subalgebra $\h_-$ the vector fields
$\mu_m$ , generating by the right action of $h_{-m}\in \h_-$ commute with
each other. So do corresponding vector fields $\eta_m$ on $\pi_0$ .
Using definition of $\eta_m$ (30) , we find commutativity of the positive AKNS
hamiltonians.
\medskip\noindent
{\bf Concluding Remarks}
\medskip\noindent
In this paper we showed the relation between classical Maxwell-Bloch
equation and AKNS hierarchy with the geometry of the affine Lie group
coset $\Y$. The simplisity of the action of the Maxwell-Bloch hamiltonian
and the integrals of motion on the phase space enables to treate
coordinates on the coset $\Y$  as "scatering data". The technique
of the Lie group interpretation of the phase space was transmited
from classical Toda theories [1-2] to the classical Maxwell-Bloch equation.
It is possible to give natural Poisson group structure on the Maxwell-Bloch
phase space
contrary to the Toda phase space. Namely, the action of the hamiltonian
on the MB phase space can be understood as a Poisson action of a Poisson-Lie
group on a Poisson manifold. Quantization of such Poisson structures leads
to a quantum version of Maxwell-Bloch equation.
Another way to quantize Maxwell-Bloch equation is to use vertex operator
algebra methods as it was done for Toda theories [1].

The lattice version of MB eq. can be given as follows. Consider functions
$$
X(z)=\beta(z) e^{\phi(z)}\,\,\hbox{ and }\,\,Y(z)=\gamma(z) e^{-\phi(z)}
$$
and introduce $X_i=X(i a)\,,\,Y_i=Y(i a), \quad i\in{\bf Z}$ as
lattice variables, where $a$ is a lattice parameter.
The Poisson bracket of $X_i$ and $Y_j$ computed using definition (1)
is a classical limit  of a q- commutator
of lattice variables $[X_i\,,\,\,Y_j]_q=\delta_{i j}$ . We will describe
the lattice version of MB eq. in the forthcoming paper.

Such kind of latticization was proposed by one of the authors (B. F) [12]
and studied for
Toda systems in [13,14].
\medskip\noindent
{\bf Acknoledgements}
\medskip\noindent
We are grateful to A.~Belavin and members of his Seminar in Landau Institute:
A.~Kadeishvili, S.~Kryukov, M.~Lashkevich, S.~Parkhomenko, V.~Postnikov
and Ya.~Pugai for useful discussions.
\medskip\noindent
{\bf References}
\medskip
1. B.~Feigin, E.~Frenkel, in Proceedings of the Summer School {\it Integrable
Systems and Quantum Groups}, Montecatini Terme, Italy, June 1993, Lect.
Notes in Math., Springer- Verlag.

2. B.~Feigin, E.~Frenkel, Generalized KdV Flows and Nilpotent Subgroups of
Affine Kac-Moody Groups, hep-th 9311171

3. M.~Ablovitz, D.~Kaup, A.~Newell, H.~Segur, {\it Stud. Appl. Math}
{\bf 53} (1974) 249-315

4. A.~Mikhailov, M.~Olshanetsky, V.~Perelomov, {\it Commun. Math. Phys.}
{\bf 79} (1981) 473-488

5. I.~Gabitov, V.~Zakharov, A.~Mikhailov {\it Teor. Mat. Fiz. (in Russian)}
{\bf 63} (1985) , 11-31

6. A.~Leznov, V.~Man'ko, S.~Chumakov {\it Teor. Mat. Fiz. (in Russian)}
{\bf 63}  (1985) , 50-63

7. L.~Faddeev, L.~Takhtajan, Hamiltonian methods in the Theory of Solitons,
Springer- Verlag, 1984

8. D.~Kazhdan, G.~Lusztig, {\it Proc. Symp. in Pure Math.} {\bf 36} (1980),
185-203

9. V.~Kac, D.~Peterson, {\it Proc. Natl. Acad. Sci. USA} {\bf 80} (1983),
1778-1782

10. A.~Pressley, G.~Segal , Loop Groups, Clarendon Press, Oxford 1986

11. J.~Bernstein, I.~Gelfand, S.~Gelfand, in Representation of Lie groups,
ed. I. Gelfand, 21-64, Wiley, New York 1975

13. B.~Feigin, Lectures in Landau Institute for Theoretical Physics,
Moscow (1992)

13. S.~Kryukov, Ya.~Pugay, Lattice W-algebras and quantum groups, hep-th
9310154

14. B.~Enriquez, B.~Feigin, Integrals of Motion of Classical Lattice
Sin-Gordon System , hep-th 9409075

\bye